\documentclass[runningheads,a4paper]{llncs}

\usepackage{amssymb}
\usepackage{graphicx}

\usepackage{url}
\urldef{\mailsa}\path|{alfred.hofmann, ursula.barth, ingrid.haas, frank.holzwarth,|
\urldef{\mailsb}\path|anna.kramer, leonie.kunz, christine.reiss, nicole.sator,|
\urldef{\mailsc}\path|erika.siebert-cole, peter.strasser, lncs}@springer.com|
\newcommand{\keywords}[1]{\par\addvspace\baselineskip
\noindent\keywordname\enspace\ignorespaces#1}

\usepackage{amssymb}
\usepackage{amsmath,bm}
\usepackage{algorithm}
\usepackage{algorithmic}
\usepackage{graphicx}

\newcounter{DefNum}
\usecounter{DefNum}
\newcounter{ExaNum}
\usecounter{ExaNum}
\renewcommand\figurename{Figure}
\begin{document}

\mainmatter  

\title{Evidential community detection based on density peaks}

\titlerunning{Evidential community detection based on density peaks}

%
%


%


%
%

\author{Kuang Zhou\inst{1} \and Quan Pan \inst{1} \and Arnaud Martin\inst{2} 
}
\authorrunning{Kuang Zhou et al.} 
%
%
\institute{Northwestern Polytechnical University,  Xi'an, Shaanxi 710072, PR China
\and
DRUID, IRISA, University of Rennes 1, Rue E. Branly, 22300 Lannion, France
\\
kzhoumath@163.com,  quanpan@nwpu.edu.cn, Arnaud.Martin@univ-rennes1.fr}

\toctitle{Lecture Notes in Computer Science}
\tocauthor{Authors' Instructions}
\maketitle

\begin{abstract}
Credal partitions in the framework of belief functions can give us a better understanding of the analyzed data set.
In order to find credal community structure in graph data sets, in
this paper, we propose a novel evidential community detection algorithm based on density peaks (EDPC).
Two new metrics,  the local density $\rho$ and the minimum dissimilarity $\delta$, are first defined for each node in the graph.
Then the nodes with both  higher $\rho$ and $\delta$ values are identified as community
centers. Finally, the remaing nodes are assigned with corresponding community
labels through a simple two-step evidential label propagation strategy. The membership of each node is described in the form
of basic belief assignments, which can  well express the uncertainty included in the community structure of the graph.
The experiments demonstrate the effectiveness of the proposed method on real-world networks.
\end{abstract}

\section{Introduction}

Community structure is one of the primary features in graphs which can gain us  a better understanding of organizations and functions in the real networked systems. As a result,
community detection, which can extract specific structures from
complex networks, has attracted considerable attention in many areas.

In 2014, Rodriguez and Laio have proposed a density peak clustering
method (DPC) in Science \cite{rodriguez2014clustering}. It is an effective and powerful
tool for the task of clustering, as neither optimization nor iteration is required in the algorithm.
DPC only provides us with a hard partition of the analyzed data set.
However, many real-world networks contain uncertain community structure, such as bridge nodes and outliers.
Credal partitions in the framework of belief functions can give us a better
understanding of the uncertain class structures of the analyzed data set.

In Ref. \cite{Zhou2017SELP}, an evidential label propagation algorithm
was introduced, where only the whole frame is used to express the uncertainty of the class structure
but the partial ignorance is not considered.
In this paper, an algorithm for detecting credal community structure, which can well describe both the
total and partial ignorance about nodes' community, is proposed based
on the concept of density peaks. Two new metrics,
the local density $\rho$ and the minimum dissimilarity $\delta$, are first defined for each node in the graph.
Then the nodes with both  higher $\rho$ and $\delta$ values can be identified as community
centers. Finally, the rest of the nodes are assigned with corresponding community
labels with a simple two-step evidential label propagation strategy.
The experiments  show that meaningful partitions of the graph could be obtained by
the proposed detection approach and it indeed could provide us more informative information of the graph structure.

The remainder of this paper is organized as follows. The density peak based clustering  is briefly introduced in Section 2.
The proposed community detection approach is presented in detail in Section 3. Some experiments on graph data sets are conducted to show the  performance in Section 4. Conclusions are drawn in the final section.

\section{Density peak based clustering}
Rodriguez and Laio \cite{rodriguez2014clustering} proposed a fast clustering approach
by finding density peaks, denoted by DPC. The idea is that
cluster centers are characterized by a higher density
than their neighbors and by a relatively large distance from any points with higher densities \cite{rodriguez2014clustering}.
From this point of view, the cluster center selection problem can be converted
into the problem of detecting outliers through a defined decision graph
using two delicately designed measures:
\begin{equation}\label{rhoi}
  \rho_i = \sum_j \chi(d_{ij}-d_c)
\end{equation}
and
\begin{equation}\label{deltai}
  \delta_i = \begin{cases}
    \max\limits_j (d_{ij}), & \text{if}~ \rho_i = \max\limits_k (\rho_k) \\
    \min\limits_{j:\rho_j >\rho_i}(d_{ij}), & \text{otherwise}
  \end{cases}
\end{equation}
The value $\rho_i$ is called the local density of point $i$. In Equation~\eqref{rhoi}, $d_{ij}$ is the distance between points $i$ and $j$,
$d_c$ is a cut-off distance. $\chi(x)$ is an indicator function which equals to 1 when $x<0$, and  0 otherwise.

The decision graph is then generated by taking $\rho_i$ as $x$ axis and
$\delta_i$ as $y$ axis. Those points with both relatively large $\rho_i$ and $\delta_i$, which are located in
the upper right corner of the graph and far away from other points, are chosen
as the centers of classes. The rest patterns  can be assigned into the same cluster as its nearest neighbor of higher density
in a single step.

\section{Evidential density-based community detection}
Inspired by the idea of density peaks, in this section we will introduce a fast evidential community detection approach based on
density peaks of graphs (denoted by EDPC).
Consider the network $G(V,E)$, where $V=\{n_1,n_2,\cdots,n_N\}$ is the set of $N$ nodes, and $E$ is the set of edges.
Denote the adjacency matrix by $\bm{A}=(a_{ij})_{N\times N}$, where $a_{ij}=1$ indicates that there is
a direct edge between nodes $n_i$ and $n_j$. Let $a_{ii}=1$.

\subsection{The dissimilarity between nodes}
In the task of community detection, the available information is often the adjacency matrix, representing the
topological structure of the graph. The similarities or dissimilarities between nodes can be determined based on the graph structure.

In this work, the dissimilarity measure based on signaling propagation process in the network is adopted, as it
can map the topological structure  into  $N$-dimensional  vectors in the
Euclidean space  \cite{hu2008community}.
For a network with $N$ nodes, every node is viewed as an excitable system which can send, receive, and record signals. Initially, a node is
selected as the source of signal. Then
the source node sends a signal to its neighbors and itself first.
Afterwards, the nodes with signals can also send signals to
their neighbors and themselves.  After a
certain $T$ time steps, the amount distribution of signals over
the nodes could be viewed as the influence of the source
node on the whole network.


Naturally, compared with nodes in other communities, the nodes of the same community have more similar
influence on the whole network. Therefore, dissimilarities between nodes could be obtained by calculating the differences between
the amount of signals they have received.
\subsection{The density peaks} 
In DPC clustering, the local density of point $i$ describes the number of points which is very close to
this pattern (with a distance to pattern $i$ smaller than $d_c$). In social networks, the person who is the center of a community may have the following characteristics: she/he
has relation with most of the members of the group; she/he may directly contact with other persons who also play an important
role in their own communities. 
Therefore, the centers of communities should be such nodes that not only with high degree, but also
with neighbors who also have high degree. Thus we can define the local degree of node $n_i$ as:
\begin{equation}
  \rho_i^{(d)} = k_i + \sum_{\{j: a_{ij}=1\}} k_j,
\end{equation}
where $k_i$ denotes the degree of node $n_i$, which can be defined as:
\begin{equation}
  k_i = \sum_{j=1}^N a_{ij}.
\end{equation}

In graphs, some bridge nodes which have connections with many groups may also have high degree centrality. In order to
distinguish these bridge nodes with the centers, we propose a new local density measure to consider both the dissimilarities
with neighbors and the centralities:
\begin{equation}
  \rho_i = \exp \left( - \frac{1}{k_i}\sum_{j:a_{ij}=1}d_{ij}^2 \right) + \rho_i^{(d)}.
\end{equation}

For some networks with fuzzy community structure, the local density measure and the minimum dissimilarities can be regularized
to distinguish cores more accurately \cite{Li2015A}: 
\begin{equation}
  \rho_i^{*} = \frac{\rho_i}{\max\limits_i \{\rho_i\}}, ~~\delta_i^{*} = \frac{\delta_i}{\max\limits_i \{\delta_i\}}.
\end{equation}

The minimum dissimilarity of nodes defined as Equation~\eqref{deltai} is adopted to measure the  degree of dispersion among center nodes.
Similar to the idea of DPC clustering,  the initial centers of the graph can be set to the nodes with high $\rho_i$ and large $\delta_i$.
Through the 2-dimensional decision graph where one dimension is $\rho_i$ and the other is $\delta_i$,
nodes that are located right upper in the decision graph are figured out as the centers.

\subsection{Allocation of other nodes}
Assume that the set of centers obtained in the last step is $V_c \subset V$.
Thus there are $c$ communities in the graph, and let the frame of discernment be $\Omega = \{\omega_1,\omega_2,\cdots,\omega_c\}$.
The credal partition defined on the power set allows to gain a deeper insight into the community structure.
The nodes located in the overlapping areas between communities will be grouped into some imprecise classes such as $\{\omega_1, \omega_2\}$,
which indicates the indistinguishability of the membership.
The outliers will be assigned to a special class $O^*$.
We use $O^*$ instead of $\Omega$ in order to distinguish
between the total ignorance class in an open world and the
imprecise class $\Omega=\{\omega_1,\omega_2, \cdots, \omega_c\}$  for overlapping nodes.
The communities of the  nodes can be determined by the label propagation process, which can be implemented as follows.

\subsubsection{Initialization}

All the center nodes are assigned with one unique community label. As there is not any uncertainty for the communities of these centers, the
Bayesian categorical mass function can be adopted to describe its membership. For example, if the center node $n_i \in V_c$ is
assigned to community $\omega_j$, we can get:
\begin{equation}
  m^i(A) = \begin{cases}
    1, & \text{if}~ A = \{\omega_j\} \\
    0, & \text{otherwise}
  \end{cases}
\end{equation}
For the rest of nodes, as there is no information about their membership at this time, the total ignorant mass function
can be used to show their membership:
\begin{equation}
  m^j(A) = \begin{cases}
    1, & \text{if}~ A = O^* \\
    0, & \text{otherwise}
  \end{cases}
\end{equation}

\subsubsection{One round expansion}

In this step, the nodes  sharing a direct link with only one center node will be first considered.
Suppose that node $n_i$ has only linked with center $n_j \in \omega_t$, and does not link with any other centers.
Similar to the principle of the label determination process in EK-NNclus\cite{denoeux2015ek}, the mass function of the node $n_i$'s  membership
can be constructed as:
\begin{equation}\label{onebba}
  m^i(A) = \begin{cases}
    \alpha, & \text{if}~ A = \{\omega_t\} \\
    1 - \alpha, & \text{if}~ A = O^\ast \\
   0, & \text{otherwise} \\
  \end{cases}
\end{equation}
where $\alpha$ is the discounting parameter such that $0 \leq \alpha \leq 1$, and it can be determined by the dissimilarity between nodes $n_i$ and $n_j$. If the dissimilarity between the two nodes
is small, that is to say, the two nodes are very close, they are most probably in the same community. Thus $\alpha$ can be set as
a decreasing function of $d_{ij}$. In this work, we suggest to use: 
\begin{equation}\label{disalpha}
  \alpha =  \exp\left\{-\gamma d_{ij}^\beta\right\},
\end{equation}
where parameter $\beta$ can be set to be 2 as default, and $\gamma$ can be set to:
\begin{equation}\label{gammaeq}
  \gamma = 1/\text{median}\left(\left\{d_{ij}^\beta,~ i=1,2,\cdots,n, ~j \in N_i\right\}\right).
\end{equation}

If one node shares a direct edge with more than one center nodes, it may be located in the overlap between/among these communities.
Suppose that node $n_i$ links with centers $n_{j_1}, n_{j_2},\cdots, n_{j_t}$, and the communities of the
$t$ centers are $\omega_{j_1},\omega_{j_2},\cdots,\omega_{j_t}$ respectively. The mass function for node $n_i$ can be defined as:
\begin{equation}
  m^i(A) = \begin{cases}
     w & \text{if}~ A = \{\omega_{j_1}, \omega_{j_2},\omega_{j_t}\} \\
      1 - w, & \text{if}~ A = O^\ast \\
     0, & \text{otherwise} \\
  \end{cases}
\end{equation}
where $w$ should be in inverse proportion to
the variation of dissimilarities between nodes $n_i$ and the corresponding centers. If the variation is small, it indicates
that there is a large amount of uncertainty for the membership of node $n_i$ and the belief assigned to the imprecise class
is large. In this paper, we use: 
\begin{equation}
  w = \exp\left\{-\mathrm{Var}(d_{ij_1}, \cdots, d_{ij_t})\right\}.
\end{equation}

\subsubsection{Diffusion in the whole network}

The unlabeled nodes will be assigned to the existing communities based on their neighbors.
The labeled nodes in the neighbors can be seen as a source of evidence. The more labeled neighbors, the more
information for the node's membership. Therefore, the update order of the unlabeled nodes should be
determined by labeled rate \cite{ding2018community}, which is defined as:
\begin{equation}
  \psi_i = \frac{|N_i^L|}{|N_i|},
\end{equation}
where $|N_i|$ denote the number of neighbors of node $n_i$, and $|N_i^L|$ denote the number of labeled neighbors.
The unlabeled node with highest $\psi_i$ are first chose for assigning a community label. Suppose that  node $n_i$ is
the one with highest labeled rate,  the
evidence provided by its $|N_i|$  neighbors are in the form of BBAs, $m^i_1, m^i_2, \cdots, m^i_{|N_i|}$,
the BBA for node $n_i$'s community membership can be obtained by combing the $N_i$ pieces of evidence from its neighbors.

The combination process can be proceeded in two steps. The first step is to divide the BBA into
different groups based on the focal element except $O^*$, and then to combine the BBAs in each group.  As there
is no conflict at all among these BBAs in the same group,
we can use the Dempster's rule directly for the inner group combination. The next step is to combine the fused BBA in different groups.
Each group can be regarded as a source for the outer  combination.
The reliability of one source  is related to the proportion of BBAs in this group. The larger
the number of BBAs in one group, the more reliable the source is.  Then the reliability discounting factor  can be defined as:
      \begin{equation}
      \label{discountfactorSimple}
      \alpha_k = \frac{s_k}{\displaystyle  \sum_{i} s_i},
      \end{equation}
where $s_k$ 
denotes the number of BBAs in each group.  The discounted BBAs in different groups are combined using
the Dubois and Prade rule~\cite{dubois1988representation} to represent the partial ignorance.
Finally, after the mass functions for all the nodes' credal membership are determined, each node can be partitioned into the  community with maximal mass assignment among all the focal elements.

\section{Experiments}

\begin{center} \begin{figure}[!htp] \centering
		\includegraphics[width=0.45\linewidth]{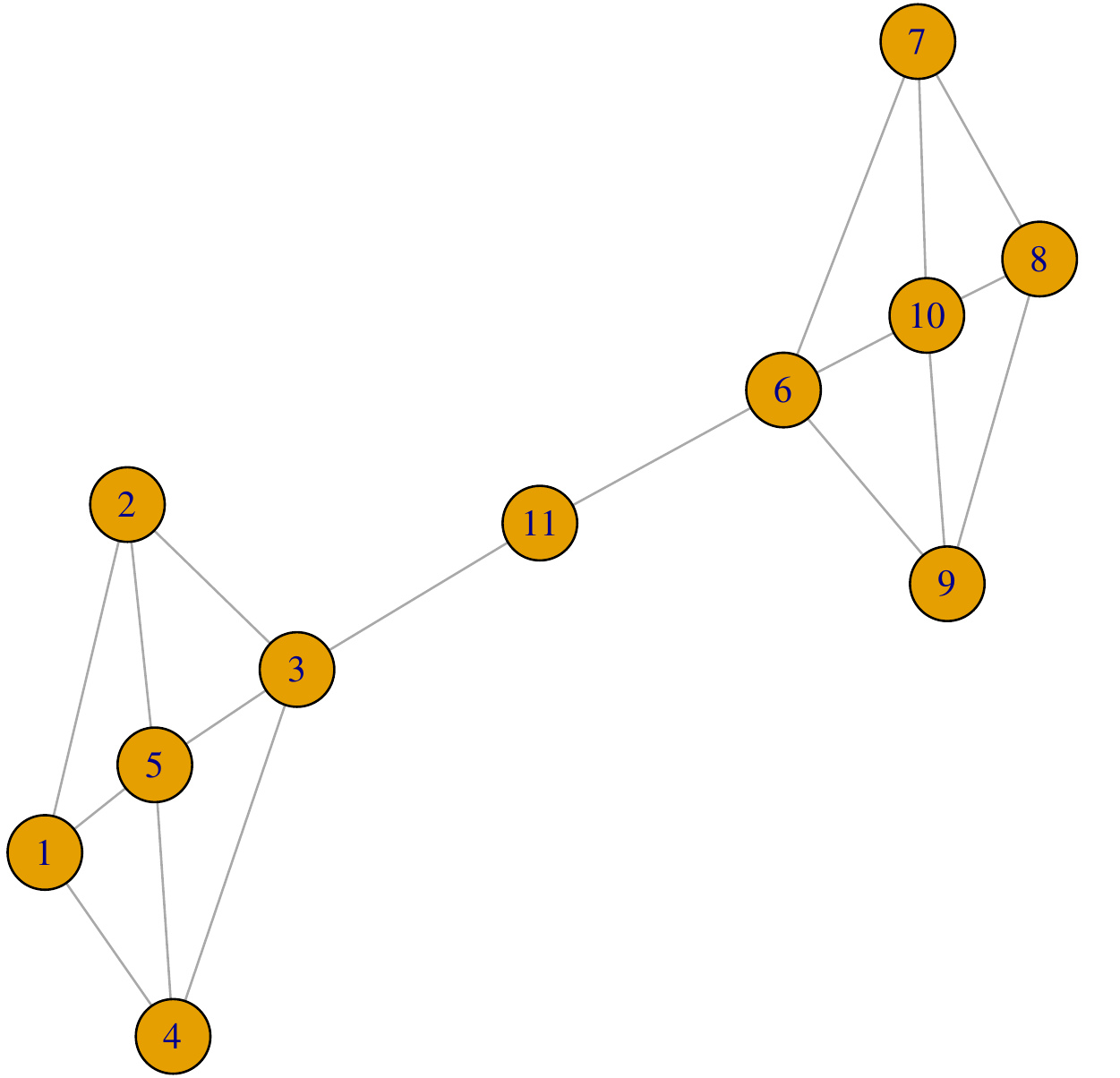}\hfill
        \includegraphics[width=0.45\linewidth]{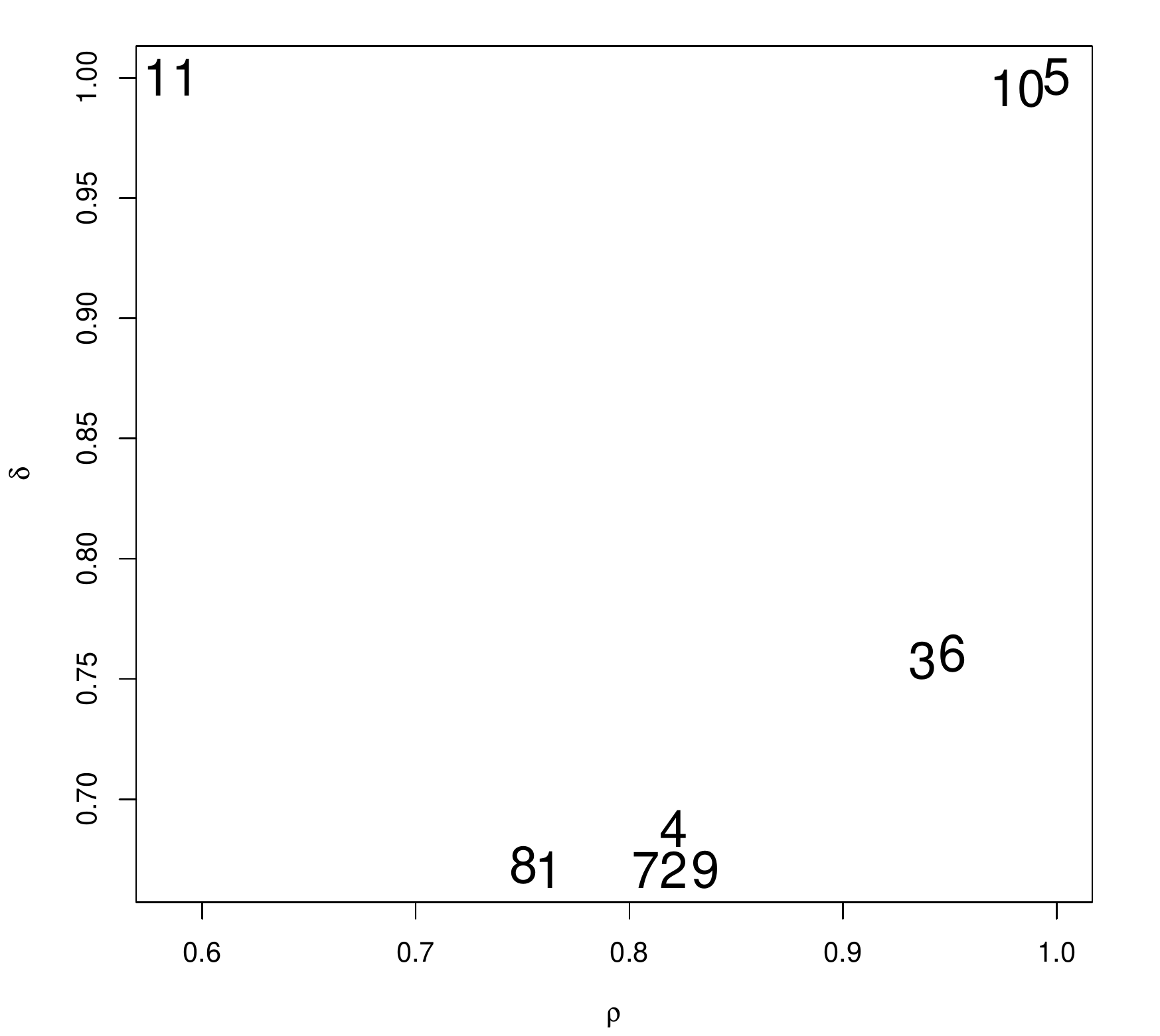} \hfill
        \parbox{.45\linewidth}{\centering\small a. The illustrative graph} \hfill
		\parbox{.45\linewidth}{\centering\small b. The decision graph}
\caption{An illustrative graph with 11 nodes.} \label{graph1} \end{figure} \end{center}

\textbf{Experiment 1.} In order to show the process of EDPC algorithm clearly, in the first experiment,
we will consider a small illustrative graph with 11 nodes displayed in Figure \ref{graph1}-a.
As can be seen from the figure, there are obviously two communities in the graph, and nodes 5 and 10 are the cores of the group, and node
11 serves as a bridge between two communities. From the decision graph in Figure \ref{graph1}-b, we can see that both center
nodes can be easily detected.

%
%

\vspace{-4em}
\begin{table}[ht]
\centering \caption{The BBAs for the 11 nodes after the first round expansion.}
\begin{tabular}{ccccccccccccc}
  \hline
 Node & $\omega_1$ & $\omega_2$ & $\Omega=\{\omega_1, \omega_2\}$ & $O^*$ \\
  \hline
1,2,3,4 & 0.6065 & 0 & 0 & 0.3935 \\
  5 & 1 & 0 & 0 & 0 \\
  6,7,8,9 & 0 & 0.6065 & 0 & 0.3935 \\
  10 & 0 & 1 & 0 & 0 \\
  11 & 0 & 0 & 0 & 1 \\
   \hline
\end{tabular}\label{graph1BBA}
\end{table}

\vspace{-2.4em}

In the first round expansion, according to the principle to determine the BBA, the membership for nodes 1, 2, 3, 4 and 6, 7, 8, 9 can be
identified using Equation~\eqref{onebba}. After the expansion, the BBA for 10 nodes in the graph
have already be determined, which can be found in Table \ref{graph1BBA}.

From this table we can see, nodes 1-4 are partitioned into the community of center node 5, while nodes 6-9 are grouped into the community of
center node 10. As node 11 has no connection with both center nodes, we have not any information for its membership after the first round
expansion. Thus the total ignorance mass function is still used to expression its membership.

In the diffusion process, the BBA for node 11 can be determined. The evidence for updating the membership of node 11 is from its neighbors,
node 3 and node 6. Using the combination rule presented in Section 3, we can get the BBA for node 11 which is listed in Table \ref{graph1BBA11}.

As can be seen from the table, node 11 is assigned with the largest belief to imprecise class $\{\omega_1, \omega_2\}$.
It reflects the indistinguishability of its membership and its bridge role between the two communities.

\begin{table}[ht]
\centering \caption{The BBA for node 11 after the diffusion.}
\begin{tabular}{ccccccccc}
  \hline
 Node & $\omega_1$ & $\omega_2$ & $\Omega=\{\omega_1, \omega_2\}$ & $O^*$ \\
  \hline
  11 & 0.2387 & 0.2387 & 0.3678 & 0.1548 \\
   \hline
\end{tabular}\label{graph1BBA11}
\end{table}

\vspace{-1em}
\noindent \textbf{Experiment 2.} To further test our proposed method, EDPC was applied to four real
networks\footnote{http://www-personal.umich.edu/~mejn/netdata/}:
Karate Club, American college football, Dolphin and Books about US politics, which have been widely used as test
networks. Two commonly used community detection methods, the label propagation algorithm (LPA), 
the modularity-based optimization method and the median evidential $c$ means clustering (MECM) based approach, 
are used for comparison. The parameters in EDPC are all set as default.
The NMI values of the obtained community structure by different methods are
reported in Table \ref{ucigraphs}. It is noted here
for EDPC, each node is partitioned into the specific community with maximal belief assignment among all the singleton focal elements.
The results show EDPC performs best in most of the data sets. It is noted that MECM based community detection method
also provides credal partitions.
The behavior of MECM and EDPC is similar, but EDPC is more efficient as it does not require iterative optimization.

\begin{table}[ht]
\centering \caption{Comparison of EDPC and other algorithms by NMI in UCI graphs.}
\begin{tabular}{cccccccccc}
  \hline
& Karate & Football & Dolphins  & Books \\
  \hline
  EDPC & \textbf{1.0000} & \textbf{0.9346} & \textbf{1.0000}  & 0.6428 \\
  MMO & 0.6873 & 0.8550 & 0.4617 & 0.5121 \\
  LPA & 0.8255 & 0.9095 & 0.8230  & 0.5485 \\
  MECM & \textbf{1} & 0.9042 & 1  & \textbf{0.7977} \\
\hline
\end{tabular}\label{ucigraphs}
\end{table}

\vspace{-3em}
\section{Conclusion}
In this paper, a novel evidential community detection approach, named EDPC, was presented inspired from the
idea of density peak based clustering. The local density of each node was defined based on its  centrality and
the dissimilarities with its neighbors. The centers were identified according to the
density and the minimum dissimilarity with the nodes with larger densities. A simple two-step evidential label propagation
strategy was designed for grouping the rest of nodes. EDPC can provide us the credal community structure of the network,
which enables us to gain a better insight into the graph structure. The experimental results have shown the effectiveness of
the proposed method.

\section*{Acknowledgements}
\vspace{-1em}
This work was supported by the National Natural Science Foundation of China (Nos.61701409, 61403310, 61672431, 61135001).

\vspace{-1em}
\bibliographystyle{splncs03}
\bibliography{paperlist}
\end{document}